
%
%
%
%
%
%
%

\magnification=\magstep1
for double spacing
\baselineskip=12pt

\font\gross=cmbx10 scaled \magstep2
\font\mittel=cmbx10 scaled\magstep1

\font\sc=cmcsc10

\font\matbf=cmmib10

\def\RR{\rm I\!R}

\def\h#1{{\cal #1}}
\def\a{\alpha}
\def\b{\beta}
\def\g{\gamma}
\def\d{\delta}

\def\l{\lambda}
\def\m{\mu}
\def\n{\nu}

\def\s{\sigma}
\def\om{\omega}
\def\na{\nabla}

\def\sq{\Square}
\def\square#1{\mathop{\mkern0.5\thinmuskip\vbox{\hrule
    \hbox{\vrule\hskip#1\vrule height#1 width 0pt\vrule}\hrule}
    \mkern0.5\thinmuskip}}
\def\Square{\mathchoice{\square{6pt}}{\square{5pt}}
    {\square{4pt}}{\square{3pt}}}

%

{\nopagenumbers
\null
\vskip-1.5cm
\vskip-.55cm
\vskip1.5mm
\vskip0.1mm
\vskip-.55cm
\vfill

\centerline{\gross New algebraic methods for calculating}
\medskip
\centerline{\gross the heat kernel and the effective action }
\medskip
\centerline{\gross in quantum gravity and gauge theories}
\bigskip
\bigskip
\centerline{{\mittel I. G. Avramidi}
\footnote{*}{ Alexander von Humboldt Fellow}
\footnote{$\S$}{On leave of absence from Research Institute for Physics,
Rostov State University, Stachki 194, Rostov-on-Don 344104, Russia}
\footnote{\dag}{Talk given on the Conference `Heat Kernel Techniques and
Quantum Gravity', University of Manitoba, Winnipeg, Canada, August 2-6, 1994}}

\centerline{\it Department of Mathematics, University of Greifswald}
\centerline{\it Jahnstr. 15a, 17489 Greifswald, Germany}
\centerline{\sl E-mail: avramidi@math-inf.uni-greifswald.d400.de}

\bigskip
\smallskip
\vfill
\centerline{\sc Abstract}
\bigskip
{\narrower
An overview about recent progress in the calculation of the heat kernel and the
one-loop effective action in quantum gravity and gauge theories is given. We
analyse the general structure of the standard Schwinger-De Witt asymptotic
expansion and discuss the applicability of that to the case of strongly curved
manifolds and strong background fields. We argue that the low-energy limit in
gauge theories and quantum gravity, when formulated in a covariant way, should
be related to background fields with covariantly constant curvature, gauge
field strength and potential term. It is shown that the condition of the
covariant constancy of the background curvatures brings into existence some Lie
algebra. The heat kernel operator for the Laplace operator is presented then as
an average over the corresponding Lie group with some nontrivial Gaussian
measure. Using this representation the heat kernel diagonal is obtained. The
result is expressed purely in terms of curvature invariants and is explicitly
covariant. Related
 topics concerning the structure of symmetric spaces and the calculation of the
effective action are discussed.

{\bigskip}}
\eject}


\centerline{\mittel 1. Introduction}
\bigskip

The heat kernel proved to be a very powerful tool both in quantum field theory
and in mathematical physics. It has been the subject of much
investigation in recent years in physical as well as in mathematical
literature
[1-22].
The study of the heat kernel is motivated, in particular, by the fact that it
gives a general framework of covariant methods for investigating the effective
action in quantum field theories with local gauge symmetries, such as quantum
gravity and gauge theories, due to special advantages achieved by using
geometric methods.

Let us consider a set of quantized (bosonic) fields $\varphi=\{\varphi^A(x)\}$
on a $d$-dimensional
Riemannian manifold $M$ without boundary ($\partial M=\emptyset$) of metric
$g_{\m\n}$ with Euclidean signature. The one-loop contribution of the field
$\varphi$ to the effective action is expressible in terms of the functional
determinant of some elliptic differential operator $F$ that can best be
presented using the $\zeta$-function regularization [1]
$$
\Gamma_{(1)}={1\over 2}\log{\rm Det}{F\over \m^2}
={1\over 2}{\rm Tr}\log{F\over \m^2}=-{1\over 2}\zeta'(0), \eqno(1.1)
$$
where $\m$ is a renormparameter introduced to preserve dimensions.
The $\zeta$-function is defined by
$$
\zeta(p)=\m^{2p} {\rm Tr}\, F^{-p} =  {\m^{2p} \over \Gamma (p)}
 \int\limits_0^\infty dt\ t^{p-1} {\rm Tr}\, U(t), \eqno(1.2)
$$
where Tr means the functional trace
$$
{\rm Tr}U(t)=\int\limits_M dx g^{1/2} {\rm tr} [U(t)],\eqno(1.3)
$$
tr is the usual matrix trace, $U(t)=\{U^A_{\ B}(t|x,x')\}$ is the corresponding
heat kernel
$$
U(t|x,x') = \exp(-t F)\h P(x,x')g^{-1/2}\d(x,x'),\eqno(1.4)
$$
where $\h P(x,x')=\{\h P^A_{\ B'}(x,x')\}$ is the parallel displacement
operator of the field $\varphi=\{\varphi^A\}$  from the point $x$  to the point
$x'$  along the geodesic, and $[U(t)]$ is the heat kernel diagonal
$$
[U(t)]=U(t|x,x).\eqno(1.5)
$$

For a very wide range of models in quantum field theory it is sufficient to
consider the second order operators of Laplace type
$$
F=-\Square +Q+m^2,\eqno(1.6)
$$
where
$$
\sq=g^{\m\n}\na_\m\na_\n=g^{-1/2}
(\partial_\m+\h A_\m)g^{1/2}g^{\m\n}(\partial_\n+\h A_\n) \eqno(1.7)
$$
is the Laplacian (or Dalambertian in hyperbolic case), $\h A_\m=\{\h A^A_{\
B\m}\}$ being an arbitrary linear connection, $Q(x)=\{Q^A_{\ B}(x)\}$ an
arbitrary  matrix-valued potential term and $m$ a mass parameter. The
connection $\h A_\m$ includes both the Levi-Civita connection, the appropriate
spin one and the vector gauge connection and is determined by the commutator of
covariant derivatives
$$
[\na_\m,\na_\n]\varphi={\cal R}_{\m\n}\varphi. \eqno(1.8)
$$
The Riemann curvature tensor $R_{\m\n\a\b}$, the curvature of background
connection ${\cal R}_{\m\n}=\{\h R^A_{\ B\m\n}\}$ and the potential term
completely describe the background metric and connection, at least {\it
locally}. In the following we will call these quantities the {\it background
curvatures} or simply curvatures and denote them symbolic by
$\Re=\{R_{\m\n\a\b}, {\cal R}_{\m\n}, Q \}$.


Let us make some remarks on the subject.
\item{$\bullet$}
First of all, it is obviously impossible to evaluate the heat kernel and the
effective action {\it exactly}, even at the one-loop order. There are, of
course, some simple special cases of background fields and geometries that
allow the exact and even explicit calculation of the heat kernel or the
effective action. However, the effective action is an {\it action}, i.e. a {\it
functional} of background fields that should be varied to get the Green
functions, the vacuum expectation values of various fields observables, such as
energy-momentum tensor and Yang-Mills currents etc.. That is why one needs the
effective action for the {\it general} background and, therefore, one has to
develop consistent {\it approximate} methods for its calculation.
\item{$\bullet$}
Second, in quantum gravity and gauge theories the effective action is a {\it
covariant} functional, i.e. invariant under diffeomorphisms and local gauge
transformations. That is why the approximations for calculating the effective
action have to preserve the general covariance at {\it each order}. The flat
space perturbation theory is an example of bad approximation because it is not
covariant.

\bigskip
\bigskip
\centerline{\mittel 2. Schwinger - De Witt asymptotic expansion}
\bigskip

First of all, let us mention the very important so called  Schwinger -  De
Witt asymptotic expansion of the heat kernel at $t\to 0$
[1-5]
$$
\eqalignno{
& {\rm Tr}\, U(t) = (4\pi t)^{-d/2}\exp(-tm^2)
\sum\limits_{k=0}^\infty {(-t)^k\over k!} B_k,        &(2.1)\cr
&B_k = \int_M dx g^{1/2} {\rm tr}b_k.                     &(2.2)\cr}
$$

This expansion is purely local  and does not depend, in fact, on the global
structure of the manifold. Its famous coefficients $b_k$ (we call them
Hadamard - Minakshisundaram - De Witt - Seely (HMDS) coefficients)
are  local invariants built from the curvature,  the potential term
and their covariant derivatives
[1,5,6,14].
They play a very important role both in physics and mathematics
and are closely connected with various sections of mathematical physics such as
spectral geometry, index theorem, Korteweg - de~Vries hierarchy etc..
[14,22]. Therefore, the calculation of HMDS-coefficients in general case of
arbitrary background is in itself of
great importance. However, it offers a complicated technical problem. Various
methods were used for calculating these
coefficients. The pioneering De Witt's method
[1] is quite simple but gets very cumbersome at higher orders. By means of it
only three first coefficients $b_0, b_1, b_2$ were calculated. The approach of
mathematicians
[5-9, 12, 14]
differs considerably from that of physicists. It is very general but also very
complicated and seems not to be well adopted to physical problems. It allowed
to compute in addition the next coefficient $b_3$ [9].

It is the general manifestly covariant technique for calculation of the HMDS -
coefficients $b_k$ that was developed in our papers [10,4]. It proved to be
very effective and allowed to calculated the next coefficient $b_4$. Moreover,
this technique allows to analyze the general structure of all HMDS-coefficients
and to calculate them in some approximation, e.g. as expansion in curvature
etc.
In the case of scalar operators the coefficient $b_4$ is also calculated in
[11].
Analytic approach was developed in
[7],  where a general expression in closed form for these coefficients
was obtained. Very good reviews of the calculation of the HMDS-coefficients
are given in recent papers
[14].

The Schwinger - De Witt expansion is good  for small $t$, viz.
$$
t\Re\ll 1, \eqno(2.3)
$$
and thereby  in  the case of {\it massive} quantized fields in {\it weak}
background fields when
$$
\Re \ll m^2, \eqno(2.4)
$$
i.e. in the case when the Compton wave length of the massive quantum field is
much smaller than the characteristic length scale of the background fields
$$
{\hbar\over mc} \ll L. \eqno(2.5)
$$
In this case the local HMDS-coefficients $b_k$ are much smaller than the
corresponding power of the mass parameter
$$
b_k \ll m^{2k}. \eqno(2.6)
$$
Therefore, one can simply integrate over $t$ to get the $1/m^2$ asymptotic
expansion of the one-loop effective action: for odd $d$
$$
\Gamma _{(1)}={1\over 2}(4\pi )^{-d/2}\pi (-1)^{d-1\over 2}
\sum\limits_{k=0}^{\infty}{m^{d-2k}\over k!\Gamma
\left({d\over 2}-k+1\right)}B_k, \eqno(2.7)
$$
and for even dimension
$$
\eqalignno{
&\Gamma_{(1)}={1\over 2}(4\pi )^{-d/2}\Biggl\{(-1)^{d/2}
\sum\limits _{k=0}^{d/2}
{m^{d-2k}\over k!\Gamma \left({d\over 2}+1-k\right)}&\cr
&\times B_k\left[\ln {m^2\over \mu^2}-
\Psi \left({d\over 2}-k+1\right) -{\matbf C} \right]
+\sum\limits_{k={d\over 2}+1}^{\infty}
{\Gamma \left(k-{d\over 2}\right)(-1)^k\over k!m^{2k-d}}B_k\Biggr\}
&(2.8)\cr}
$$
where $\Psi(q)=(d/dq)\log\Gamma(q), \ {\matbf C}=-\Psi(1)$.

This approximation describes good the physical effect of the vacuum
polarization effect of massive quantized field in weak background fields.

\bigskip
\bigskip
{\vbox{
\centerline{\mittel 3. General structure of the asymptotic expansion}
\smallskip
\centerline{\mittel  and partial summation}
\bigskip}}
\nobreak
However the Schwinger - De Witt approximation is of very limited applicability.
It is absolutely inadequate for large $t$  ($t\Re \gg 1$), in strongly curved
manifolds and strong background fields ($\Re \gg m^2$) and becomes meaningless
in massless theories. Therefore, this approximation can not describe
essentially nonperturbative effects such as particle creation and the vacuum
polarization by strong background fields, which are, in fact, {\it nonlocal}
and {\it nonanalytical}.

In fact, the effective action is a nonlocal and nonanalytical functional and
possesses a sensible massless limit. But its calculation requires quite
different methods. One of such methods which would exceed the limits of the
Schwinger - De Witt asymptotic expansion is the {\it partial summation}
procedure [2]. It is based on the analysis of the general structure of the
HMDS-coefficients. The HMDS-coefficients are the local polynomial invariants
built from the curvature and its covariant derivatives. The first two
HMDS-coefficients have the well known form
$$
\eqalignno{
B_0 &=\int dx g^{1/2}{\rm tr} 1,            &\cr
B_1 &=\int dx g^{1/2}{\rm tr}\left\{Q-{{1}\over {6}} R\right\}      &(3.1)\cr}
$$

One can classify all the terms in the higher order coefficients $B_k$, $(k\ge
2)$, according to the number of curvatures and their derivatives. The terms
with leading derivatives can be shown to have the following structure $\Re
\sq^{k-2} \Re$. Then it follows the class of terms cubic in the curvatures
etc.. The last class of terms does not contain any covariant derivatives at all
but only the powers of the curvature
$$
\eqalignno{
B_k &= \int dx g^{1/2}{\rm tr}\,\Biggl\{\Re \sq^{k-2} \Re
+\sum_{0\le i\le 2k-6}\Re\ \na^i\Re\ \na^{2k-6-i}\Re &\cr
&+\cdots+\sum_{0\le i\le k-3}\Re^i(\na \Re)\Re^{k-i-3}(\na\Re)
+\Re^k\Biggr\}&(3.2)\cr}
$$

Now one can try to sum up each class of terms separately to get the
corresponding expansion of the heat kernel
$$
\eqalignno{
{\rm Tr}U(t)=&\int dx g^{1/2}(4\pi t)^{-d/2}\exp(-tm^2){\rm tr}
\Biggl\{1-t\left(Q-{1\over 6}R\right)		&\cr
&+t^2\Re\chi(t\sq)\Re+\cdots+
t^3\na \Re\Psi(t\Re)\na \Re + \Phi(t\Re)\Biggr\}	&(3.3)\cr}
$$
and that of the effective action
$$
\Gamma_{(1)}=\int dx g^{1/2}{\rm tr}
\Biggl\{\Re F(\sq)\Re+\cdots+\na \Re Z(\Re)\na \Re + V(\Re)\Biggr\}
\eqno(3.4)
$$

The idea of partial summation consists in comparing the values of all terms in
HMDS-coefficients $B_k$
(3.2),  picking up the main (the largest in some approximation) terms and
summing up the corresponding partial sum. There is always a lack of uniqueness
concerned with the global structure of the manifold, when doing so.  But,
hopefully, fixing the topology, e.g. the trivial one, one can obtain a unique,
well defined, expression that would reproduce the Schwinger -De Witt
expansion, being expanded in curvature.

One should mention that these expansions are {\it asymptotic} ones and {\it do
not converge}, in general. So, one has to use some methods of summation of the
divergent asymptotic series. This can be done by using an integral transform
and analytic continuation [4].

Actually, the effective action is a covariant functional of  the metric
and depends on  the
geometry of the  manifold as a whole,  i.e.  it depends on both local
characteristics of the geometry like invariants of the curvature tensor and
its global topological structure. However, we are {\it not} going to {\it
investigate} the influence of the {\it topology} but
concentrate our attention, as a rule, on the {\it local} effects. Then the
possible approximations for evaluating the effective action can be based on
the assumptions about the local behavior of the background fields, dealing
with the real physical gauge invariant variations of  the local geometry,
i.e. with the curvature invariants, but not with the behavior of the
metric and the connection which is not invariant.  Comparing the value of the
curvature with the values of its covariant derivatives one comes to two
possible approximations:
i) the {\it short-wave} (or {\it high-energy}) approximation characterized by
$$
\na\na\Re\gg\Re\Re \eqno(3.5)
$$
and ii) the {\it long-wave} (or {\it low-energy}) one
$$
\na\na\Re\ll\Re\Re. \eqno(3.6)
$$
Such a formulation is manifestly covariant and, therefore, more suitable for
calculations in quantum gravity and gauge theories than the usual flat space
perturbation theory.

\bigskip
\bigskip
\centerline{\mittel 4. High-energy approximation}
\bigskip

\def\nohkf {
    + {{t^2}\over {2}}\biggl[Q\gamma^{(1)}(t\Square)Q+
     2\h R_{\alpha\mu}\nabla^\alpha{{1}\over {\Square}}\gamma^{(2)}
     (t\Square)\nabla_\nu\h R^{\nu\mu}
          -2Q\gamma^{(3)}(t\Square)R  }
\def\nohks {+R_{\mu\nu}\gamma^{(4)}
      (t\Square)R^{\mu\nu}
              +R\gamma^{(5)}(t\Square)R\biggr]  }


The idea of partial summation was realized in short-wave approximation
for investigating the nonlocal aspects of the effective action (in
other words the high-energy limit of that) in
[15,4].
The leading terms with higher derivatives in HMDS-coefficients (quadratic in
curvatures) have the following form
$$
\eqalignno{
&B_k=\int dx\,g^{1/2}{\rm tr}\biggl\{
f^{(1)}_k Q\sq^{k-2}Q
+2f^{(2)}_k \h R_{\alpha\mu}\nabla^\alpha\sq^{k-3}\nabla_\nu\h R^{\nu\mu}&\cr
&-2f^{(3)}_k Q\sq^{k-2}R +f^{(4)}_k R_{\mu\nu}\sq^{k-2}R^{\mu\nu}+f^{(5)}_k
R\sq^{k-2}R
+O(R^3)\biggr\}			&  (4.1)\cr}
$$
The coefficients $f^{(i)}_k$ were calculated explicitly in [15,4] and
completely independent in [16].

Moreover, the local Schwinger - De Witt expansion can now be summed up to get
the nonlocal heat kernel
$$
     \eqalignno{
  {\rm Tr}\,U(t) &=
     \int dx\, g^{1/2}(4\pi t)^{-d/2}\exp{(-tm^2)}{\rm tr}\biggl\{
                1-t\left(Q-{{1}\over {6}}R\right)               & \cr
                &\nohkf                                    &\cr
                &\nohks             +O(R^3)\biggr\}. &(4.2)\cr}
$$
and the corresponding high-energy effective action
$$
\Gamma_{(1)}=\Gamma_{(1)loc}+\Gamma_{(1)nonloc}
\eqno(4.3)
$$
where the essential nonlocal part has the form
$$
       \eqalignno{
       \Gamma_{(1)nonloc}&={1\over 2}(4\pi )^{-d/2}\int dx \,g^{1/2}\,{\rm tr}
                           \biggl\{Q\beta^{(1)}(\Square)Q
                           +2\h R_{\alpha\mu}\nabla^\alpha{{1}\over {\Square}}
                           \beta^{(2)}(\Square)\nabla_\nu\h R^{\nu\mu}& \cr
                          &-2Q\beta^{(3)}(\Square)R
                          +R_{\mu\nu}\beta^{(4)}(\Square)R^{\mu\nu}
                          +R\beta^{(5)}(\Square)R
                          +O(R^3)\biggr\}                   &(4.4)\cr}
$$
The explicit form of the formfactors $\g^{(i)}(t\sq)$ and $\b^{(i)}(\sq)$ is
given in [15,4].

Another approach to study the high-energy limit of the effective action, so
called {\it covariant perturbation theory}, is
developed in
[17]. By means of it the next {\it third} order in curvatures of the nonlocal
high-energy limit is investigated.

\bigskip
\bigskip
\centerline{\mittel 5. Low-energy approximation}
\bigskip

The low-energy  effective action, in other words, the effective potential,
presents a very natural tool for investigating the vacuum of the theory, its
stability and the phase structure
[23]. In the case of gauge theories and quantum gravity it is much more
complicated problem as the high-energy limit. Here only partial success is
achieved and various approaches to this
problem are only outlined (see, e.g. our recent papers [20,21]).

The long-wave (or low-energy) approximation is determined, as it was already
stressed above, by strong slowly varying  background fields.  This means that
the derivatives of all invariants are much smaller than the  products of the
invariants themselves. The zeroth order of this approximation corresponds to
covariantly constant background curvatures
$$
\na_\m R_{\a\b\g\d} = 0,\qquad \na_\m{\cal R}_{\a\b}=0,\qquad \na_\m Q = 0.
							\eqno(5.1)
$$

In this case the HMDS-coefficients are simply polynomials in curvature
invariants and potential term of dimension $\Re^k$ up to terms with one
or more covariant derivatives of the background curvatures $O(\na \Re)$
$$
\eqalignno{
&b_k=\sum^k_{n=0}{k\choose n}Q^{k-n}a_n + O(\na \Re) ,   &(5.2)\cr
&a_k=b_k\Big\vert_{Q = \na R = 0} = \sum \Re^k    .        &(5.3)\cr}
$$
Mention that the commutators $[Q,{\cal R}_{\m\n}]$ are of order $O(\na\na \Re)$
and, therefore are neglected here.

Then after summing the Schwinger-De Witt expansion
we obtain for the heat
kernel, the $\zeta$-function and the effective action
$$
\eqalignno{
{\rm Tr}\,U(t)=\int_M dx\,g^{1/2}
&(4\pi t)^{-d/2}{\rm tr}\left\{\exp{\left(-t(m^2+Q)\right)}\left(\Omega(t)
+O(\na \Re)\right)\right\},&(5.4)\cr
\zeta(p)=\int_M dx\,g^{1/2}
&(4\pi)^{-d/2}{\m^{2p}\over\Gamma(p)}\int\limits_0^\infty dt
\ t^{p-d/2-1}&\cr
&\times{\rm tr}\left\{\exp{\left(-t(m^2+Q)\right)}\left(\Omega(t)
+O(\na\Re)\right)\right\},&(5.5)\cr
\Gamma_{(1)}=\int_Mdx\,g^{1/2}&\{V(\Re)
+O(\na\Re)\},&(5.6)\cr}
$$
with
$$
\eqalignno{
V(\Re) = {1\over 2}(4\pi)^{-d/2}&{1\over \Gamma({d\over 2}+1)}
\int\limits_0^\infty dt \left(\log (\m^2 t)+\psi\left({d\over
2}+1\right)\right) &\cr
&\times\left({\partial \over \partial t}\right)^{{d\over 2}+1}{\rm
tr}\left\{\exp(-t(m^2+Q))\Omega(t)\right\} &(5.7)\cr}
$$
for even $d$ and
$$
\eqalignno{
V(\Re) &= {1\over 2}(4\pi)^{-d/2}{1\over \Gamma\left({d\over 2}+1\right)}&\cr
&\times\int\limits_0^\infty dt t^{-1/2}\left({\partial \over \partial
t}\right)^{{d+1\over 2}}{\rm tr}\left\{\exp(-t(m^2+Q))\Omega(t)\right\}
&(5.8)\cr}
$$
for odd $d$, where
$$
\Omega(t) =  \sum\limits_{k=0}^\infty {(-t)^k\over k!} a_k  , \eqno(5.9)
$$
is a function of local invariants of the curvatures (but not of the potential).

It is naturally to call the functions $\Omega(t)$ and $V(\Re)$,  that do not
contain the {\it covariant} derivatives at all and so determine the zeroth
order of the heat kernel and that of the effective action, the {\it
generating function} for covariantly constant terms in HMDS-coefficients and
the {\it
effective potential} in quantum gravity respectively.

Let us mention that such a definition of the effective potential is not
conventional. It differs from the definition that is often found in the
literature
[24]. What is meant usually under the notion of the effective potential
is a function of the potential term only $Q$, because it does not  contain
derivatives of the background field (in contrast to Riemann curvature
$R_{\a\b\g\d}$ that contains second derivatives of the metric and the curvature
${\cal R}_{\m\n}$ with first derivatives of the connection).  So, e.g. in
[24] the  potential term $Q$ is summed up exactly but an expansion is
made not only in covariant derivatives but also in powers of curvatures
$R_{\m\n\a\b}$ and ${\cal R}_{\m\n}$, i.e. the curvatures are treated
perturbatively. Thereby the
validity of this approximation for the effective action  is limited to  small
curvatures
$$
{\cal R}_{\m\n}, R_{\m\n\a\b} \ll Q. \eqno(5.9a)
$$
Such an expansion is called {\it `expansion of
the effective action in covariant derivatives'.} Without the potential term
($Q=0$) the effective potential in such a scheme is trivial. Hence we stress
here once again, that the effective potential in our definition contains, in
fact,  much more information than the usual effective potential does
when using the `expansion in covariant derivatives'.
As a matter of fact, what we mean is the {\it low-energy limit of the
effective action} formulated in a covariant way.

The conditions of integrability of these relations lead to  strong
algebraic restrictions on the curvatures themselves
$$
\eqalignno{
&R_{\m\n\l[\a}R^\l_{\ \b]\g\d} + R_{\m\n\l[\g}R^\l_{\ \d]\a\b} = 0&(5.10)\cr
&R_{\m\n\l[\a}\h R^\l_{\ \b]} + R_{\a\b\l[\m}\h R^\l_{\ \n]} = 0&(5.11)\cr
&[\h R_{\m\n},\h R_{\a\b}] + R_{\m\n\l[\a}\h R^\l_{\ \b]}
- R_{\a\b\l[\m}\h R^\l_{\ \n]} = 0&(5.12)\cr
&[\h R_{\m\n},Q] = 0&(5.13)\cr}
$$
Mention that the conditions (5.1), (5.10)-(5.13) are local. They determine the
geometry of the {\it locally} symmetric spaces.  However, the manifold is {\it
globally} symmetric one only in the case when it satisfies additionally some
global topological restrictions (usually it has to be connected)
and the condition (5.1) is valid everywhere, i.e. at any point of
the manifold
[25].

But in our case, i.e. in {\it physical } problems, the situation is radically
different. The correct setting of the problem seems to be as follows.
The low-energy  effective action depends, in general, also essentially on the
global topological properties of the space-time manifold.  But, as it was
mentioned above, we  do not  investigate  in this paper the influence  of the
 topology.  Therefore, consider a complete noncompact asymptotically flat
manifold without boundary that is homeomorphic to $\RR^d$.  Let  a  finite
not small, in general, domain of the manifold exists that is strongly curved
and quasi-homogeneous, i.e. the invariants of the curvature in this region
vary very slowly.  Then the geometry of this region is locally very similar
to that of a symmetric space.  However one should have in mind that there are
{\it always} regions in the manifold where this condition is not fulfilled.
This is, first of all, the asymptotic Euclidean region that has small
curvature and, therefore, the opposite short-wave approximation is valid.

The general situation in correct setting of the problem is the
following.  From infinity with small curvature and possibly radiation, where
[17] $\Re \Re \ll \na\na \Re $,  we pass on to quasi-homogeneous region where
the local properties of the manifold are close to those of symmetric spaces.
The size of this region can tend to zero. Then the curvature is nowhere large
and the short-wave approximation is valid anywhere.
If one tries to extend the limits of such region to infinity, then one has
also to analyze the topological properties. The space can be compact or
noncompact depending on the sign of the  curvature.  But first we will come
across a coordinate horizon-like singularity, although no one true physical
singularity really exists.

This construction can be  intuitively imagined as follows. Take the flat
Euclidean space $\RR^d$, cut out from it a region $M$ with some boundary
$\partial M$ and stick to it along the boundary, instead of the piece cut
out, a piece of a curved symmetric space with the same boundary $\partial M$.
Such a construction will be homeomorphic to the initial space and at the same
time will contain a finite highly curved homogeneous region. By the way,
the exact effective action for a symmetric space differs from the  effective
action for built construction by a purely topological  contribution.
This fact seems to be useful when analyzing the effects of topology.

Thus the problem is to calculate  the low-energy effective action (the
effective potential $V(\Re)$) (5.7), (5.8), i.e. the heat kernel for
covariantly constant background. Although this
quantity, generally speaking, depends essentially on the topology and other
global aspects of the manifold, one can disengage oneself  from these effects
fixing the trivial topology. Since the asymptotic Schwinger - De Witt
expansion does not depend on the topology, one can hold that we thereby
{\it sum up all the terms without covariant derivatives} in it.

In other words the problem is  the following. One has to obtain a local
covariant function of the invariants of the curvature $\Omega(t)$ (5.9) that
would describe adequately the low-energy limit of the heat
kernel diagonal and that would, being expanded in curvatures, reproduce {\it
all terms
without covariant derivatives} in the asymptotic expansion of heat kernel,
i.e. the HMDS-coefficients $a_k$ (5.3). If one finds such an expression, then
one can simply determine the $\zeta$-function (5.5) and, therefore, the
effective potential $V(\Re)$ (5.7), (5.8).


\bigskip
\bigskip
\centerline{\mittel 6. Algebraic approach}
\bigskip

There exist a very elegant indirect possibility to construct the heat kernel
{\it without solving the heat equation} but {\it using only the commutation
relations} of some {\it covariant} first order differential operators. The main
idea is in a generalization of the usual Fourier transform to the case of
operators and consists in the following.

Let us consider for a moment a trivial case of vanishing curvatures but not the
potential term
$$
R_{\a\b\g\d} = 0,\qquad {\cal R}_{\a\b}=0,\qquad Q \ne 0.
							\eqno(6.1)
$$
In this case the operators of covariant derivatives obviously commute and form
together with the potential term an Abelian algebra
$$
[\na_\m,\na_\n]=0,\qquad [\na_\m, Q]=0. \eqno(6.2)
$$

It is easy to show that the heat kernel {\it operator} can be presented in the
form
$$
\exp(t\sq)=(4\pi t)^{-d/2}\int dk g^{1/2}
\exp\left\{-{1\over 4t}k^\m g_{\m\n}k^\n
+k^\m \na_\m\right\}. \eqno(6.3)
$$
Here, of course, it is assumed that the covariant derivatives commute also with
the metric
$$
[\na_\m,g_{\a\b}]=0. \eqno(6.4)
$$
Acting with this operator on the $\d$-function and using the obvious relation
$$
\exp(k^\m \na_\m)g^{-1/2}\d(x,x')=g^{-1/2}\d(x+k,x')
\eqno(6.5)
$$
one can simply integrate over $k$ to obtain the heat kernel in coordinate
representation
$$
U(t|x,x')=(4\pi t)^{-d/2}\exp\left\{-t(m^2+Q)
-{1\over 4t}(x-x')^\m g_{\m\n}(x-x')^\n\right\}. \eqno(6.6)
$$
The heat kernel diagonal is given then by
$$
[U(t)]=(4\pi t)^{-d/2}\exp\left\{-t(m^2+Q)\right\}, \eqno(6.7)
$$
and the function $\Omega(t)$ (5.9) is simply
$$
\Omega(t)=1. \eqno(6.7a)
$$

In fact, the covariant derivatives do not commute and the commutators of them
are proportional to the curvatures. Thus one can try to generalize the above
idea in such a way that (6.3) would be the zeroth approximation in the
commutators of the covariant derivatives, i.e. in the curvatures. We are going
to find a representation of the heat kernel {\it operator} in the form
$$
\exp(t\sq) = \int dk \Phi(t,k)\exp\left\{-{1\over 4t}k^A\Psi_{AB}(t)k^B
+k^AX_A\right\}
\eqno(6.8)
$$
where $X_A=X^\m_A\na_\m$ are some first order differential operators. The
commutators of them should be proportional to the curvature
$$
[X_A,X_B]=O(\Re) \eqno(6.9)
$$
and the functions $\Psi(t)$ and $\Phi(t,k)$ should have the following property
$$
\Psi(t)=\g_{AB}+O(\Re), \qquad
\Phi(t,k)=\g^{1/2}(1+O(\Re)), \eqno(6.9)
$$
where $\g_{AB}$ is some constant nondegenerate positive definite matrix,
$\g^{1/2}=\det \g_{AB}$ and $O(\Re)$ means the terms linear in the curvatures.

In general, the operators $X_A$ do not form a closed finite dimensional algebra
because at each stage taking more commutators there appear more and more
derivatives of the curvatures. It is the case of {\it covariantly constant
curvatures} that actually {\it closes} the algebra. In this case the operators
$X_A$ together with the curvatures form some Lie algebra.

Using this representation one could, as above, act with $\exp(k_AX^A)$ on the
$\d$-function on $M$ to get the heat kernel. The main point of this idea is
that it is much more easier to calculate the action of the exponential of the
{\it first} order operator $k^AX_A$ on the $\d$-function than that of the
exponential of the second order operator $\sq$.

\bigskip
\bigskip
\centerline{\mittel 7. Heat kernel in flat space}
\smallskip
\centerline{\mittel with nonvanishing Yang-Mills background}
\bigskip

This idea was realized in [21] for more complicated case of vanishing Riemann
curvature (flat space) but nonvanishing curvature of background connection
(Yang-Mills case)
$$
R_{\a\b\g\d} = 0,\qquad {\cal R}_{\a\b}\ne 0,\qquad Q \ne 0.
							\eqno(7.1)
$$

Using the condition of covariant constancy of the curvatures (5.1),
(5.12)-(5.13) one can show that in this case the covariant derivatives form a
{\it nilpotent} Lie algebra
$$
\eqalignno{
&[\na_\m,\na_\n]=\h R_{\m\n}, &(7.2)\cr
&[\na_\m,\h R_{\a\b}]=[\na_\m,Q]=0, &\cr
&[\h R_{\m\n},\h R_{\a\b}]=[\h R_{\m\n},Q]=0.\cr}
$$

For this algebra one can prove a theorem expressing the heat kernel operator in
terms of an average over the corresponding Lie group
$$
\eqalignno{
\exp(t\sq) =& (4\pi t)^{-d/2}
\det\left({t\h R\over \sinh(t\h R)}\right)^{1/2}&\cr
&\int dk g^{1/2}
\exp\left\{-{1\over 4t}k^\m g_{\m\l}(t\h R \coth(t\h R))^\l_{\ \n} k^\n +
k^\m \na_\m\right\}				&(7.3)\cr}
$$
where $\h R$ means the matrix with coordinate indices $\h R=\{\h R^\m_{\
\n}=g^{\m\l}\h R_{\l\n}\}$  and the determinant is taken with respect to these
indices, other (bundle) indices being intact.
The proof of this theorem is given in [21].

Using this theorem we express the heat kernel in coordinate representation in
terms of the quantity
$$
\exp(k^\m \na_\m)\h P(x,x')g^{-1/2}\d(x,x')
\eqno(7.4)
$$
according to the definition of the heat kernel (1.4). It is not difficult to
show that
$$
\exp(k^\m \na_\m)\h P(x,x')g^{-1/2}\d(x,x')
=\h P(x,x')g^{-1/2}\d(x+k,x').
\eqno(7.5)
$$
Subsequently, the integral over $k^\m$ becomes trivial and one obtains
immediately the heat kernel in coordinate representation
$$
\eqalignno{
U(&t|x,x') = (4\pi t)^{-d/2}\h P(x,x')
\det\left({t\h R \over \sinh(t\h R)}\right)^{1/2}&\cr
&\times \exp\left\{-t(m^2 + Q) -
{1\over 4t}(x-x')^\m g_{\m\l}(t\h R\coth(t\h R))^\l_{\ \n}
(x-x')^\n \right\}&(7.6)\cr}
$$

The heat kernel diagonal is now easily obtained by taking the coincidence limit
$x=x'$
$$
[U(t)]=(4\pi t)^{-d/2}
\det\left({t\h R \over \sinh(t\h R)}\right)^{1/2}
\exp\left\{-t(m^2+Q)\right\}. \eqno(7.7)
$$
The generating function $\Omega(t)$ is now
$$
\Omega(t)=\det\left({t\h R \over \sinh(t\h R)}\right)^{1/2}. \eqno(7.8)
$$
It defines the $\zeta$-function (5.5) and the corresponding effective potential
(5.7), (5.8). Expanding it in a power series in $t$ one can find {\it all}
covariantly constant terms in {\it all} HMDS-coefficients $a_k$.

As we have seen the contribution of the bundle curvature $\h R_{\m\n}$ is not
as trivial as that of the potential term. However, the algebraic approach does
work in this case too. This is the generalization of the well known Schwinger
result in quantum electrodynamics when the bundle curvature is just the
electromagnetic field strength $\h R_{\m\n}=iF_{\m\n}$. It is a good example
how one can get the heat kernel without solving any differential equations but
using only the algebraic properties of the covariant derivatives.


\bigskip
\bigskip
\centerline{\mittel 8. Heat kernel in symmetric spaces}
\bigskip

Let us now try to generalize this construction to the case of the {\it curved}
manifolds with covariantly constant curvature, i.e. symmetric spaces.
Below we follow mainly our papers [20]. Let us list shortly some known facts
about symmetric spaces presented in the form that will be convenient for
further use. What are the direct consequences of the condition of covariant
constancy of the curvature?

First of all, let $e^\m_a(x,x')$ be a frame that is {\it covariantly constant
(parallel)} along the geodesic between points $x$ and $x'$. The frame
components of all tensors will be always understood with respect to this
special frame. Let us consider the Riemann tensor in more detail. It is obvious
that the frame components of the curvature tensor of a symmetric space are
constant. For {\it any} Riemannian manifold they can be presented in the form
$$
R_{abcd} = \b_{ik}E^i_{\ ab}E^k_{\ cd}, \eqno(8.1)
$$
where $E^i_{ab}$, $(i=1,\dots, p; p \le d(d-1)/2)$, is some set of
antisymmetric matrices and $\b_{ik}$ is some symmetric nondegenerate $p\times
p$ matrix.
The traceless matrices $D_i=\{D^a_{\ ib}\}$ defined by
$$
D^a_{\ ib}=-\b_{ik}E^k_{\ cb}g^{ca}= - D^a_{\ bi} \eqno(8.2)
$$
are known to be the generators of the {\it isotropy algebra} ${\cal H}$ of
dimension ${\rm dim} \h H=p$
$$
[D_i, D_k] = F^j_{\ ik} D_j, \eqno(8.3)
$$
with $F^j_{\ ik}$ being the structure constants. The structure constants
$F^j_{\ ik}$ are completely determined by these commutation relations and
satisfy Jacobi identities
$$
[F_i, F_k] = F^j_{\ ik} F_j, \eqno(8.4)
$$
where $F_i=\{F^k_{\ il}\}$ are the generators of the isotropy algebra in
adjoint representation. Mention that the isotropy group $H$ is always compact
as it is a subgroup of the orthogonal group (in Euclidean case).

It is not difficult to show that the condition of integrability (5.10) of the
equations (5.1)
takes the form
$$
E^i_{\ a c} D^c_{\ b k} -  E^i_{\ b c} D^c_{\ a k}= E^j_{\ a b} F^i_{\ j k}.
\eqno(8.5)
$$
This equation takes place {\it only} in symmetric spaces and is the most
important one. It is this equation that makes a Riemannian manifold the
symmetric space. From the eqs. (8.3) and (8.4) we have, in particular,
$$
\b_{ik} F^k_{\ jm} + \b_{mk} F^k_{\ ji} = 0, \qquad \eqno(8.6)
$$
that means that the adjoint and coadjoint representations of the isotropy group
are equivalent and the matrix $\b_{ik}$ plays the role of the metric of the
isotropy algebra.

Moreover, the eq. (8.5) brings into existence a much wider algebra ${\cal G}$
of dimension ${\rm dim}\,{\cal G}=D=p+d$. Indeed, let us define the quantities
$C^A_{\ BC}=-C^A_{\ CB}$, $(A=1,\dots, D)$ by
$$
C^i_{\ ab}=E^i_{\ ab}, \quad C^a_{\ ib}=D^a_{\ ib}, \quad C^i_{\ kl}=F^i_{\
kl}, \eqno(8.7)
$$
$$
C^a_{\ bc}=C^i_{\ ka}=C^a_{\ ik}=0,
$$
and the matrices $C_A=\{C^B_{\ AC}\}=(C_a,C_i)$,
$$
C_a = \left( \matrix{ 0          & D^b_{\ ai}   \cr
		      E^j_{\ ac} & 0            \cr}\right) ,\qquad
C_i = \left( \matrix{ D^b_{\ ia} & 0            \cr
		      0          & F^j_{\ ik}   \cr}\right) .
						\eqno(8.8)
$$

Using the eqs. (8.3)-(8.6) one can show that they satisfy the Jacobi identities
$$
[C_A, C_B]=C^C_{\ AB}C_C               \eqno(8.9)
$$
or, more precisely,
$$
\eqalignno{
[C_a, C_b] &= E^i_{\ a b} C_i, \qquad &\cr
[C_a, C_i] &= D^b_{\  a i} C_b, \qquad &(8.10)\cr
[C_i, C_k] &= F^j_{\ i k} C_j, & \cr}
$$
and define, therefore, some Lie algebra ${\cal G}$ with the structure constants
$C^A_{\ BC}$ $=$ $\{E^i_{\ a b},$ $D^b_{\ ia},$ $F^j_{\ ik}\}$, matrices $C_A$
being the generators in adjoint representation.

Further, introducing a symmetric nondegenerate $D\times D$ matrix
$$
\g_{AB} = \left(\matrix{ g_{ab} & 0             \cr
	       0                & \b_{ik}        \cr}\right), \eqno(8.11)
$$
that plays the role of the metric on the algebra ${\cal G}$ one can show that
the structure constants satisfy also the identity
$$
\g_{AB} C^B_{\ CD} + \g_{DB} C^B_{\ CA} = 0, \eqno(8.12)
$$
that means that the adjoint and coadjoint representations of the algebra ${\cal
G}$ are also equivalent.

In other words, the Jacobi identities (8.9) are equivalent to the
identities (5.10) that the curvature must satisfy in the symmetric space.
This means that the set of the structure constants $C^A_{\ BC}$, satisfying the
Jacobi identities, determines the curvature tensor of symmetric space $R^a_{\
bcd}$. Vice versa the structure of the algebra ${\cal G}$ is completely
determined by the curvature tensor of symmetric space.

Now consider the bundle curvature ${\cal R}_{ab}$. One can show analogously
that because of the integrability conditions (5.11), (5.12) it must have the
form
$$
{\cal R}_{ab}={\cal R}_i E^i_{\ ab}, \eqno(8.13)
$$
where $E^i_{\ ab}$ are the same 2-forms and ${\cal R}_i$ are some matrices
forming a representation of the isotropy algebra
$$
[{\cal R}_i, {\cal R}_k]= F^j_{\ ik}{\cal R}_j. \eqno(8.14)
$$

Finally, the potential term should commute with the curvature ${\cal R}_{\m\n}$
and, therefore, with all matrices ${\cal R}_i$
$$
[{\cal R}_i, Q]=0. \eqno(8.15)
$$

One can show that the algebra $\h G$ is isomorphic to the algebra of the
infinitesimal isometries of symmetric space. The set of all generators of
infinitesimal isometries $\h G=\{\xi_A\}, {\rm dim} \h G=D,$ can be split in
two essentially different sets: $\h M = \{P_a\}, \dim \h M=d$, and $\h H
=\{L_i\}, \dim \h H=p$,
according to the values of their initial parameters
$$
P_a\Big\vert_{x=x'}\ne 0, \qquad L_i\Big\vert_{x=x'}=0. \eqno(8.16)
$$
Let us introduce a two-point matrix $K=\{K^a_{\ b}(x,x')\}$
$$
K^a_{\ b}= R^a_{\ cbd}\s^c\s^d, \eqno(8.17)
$$
where $\s^a$ are the frame components of the tangent vector to the geodesic
connecting the points $x$ and $x'$ at the point $x'$, $\s^a(x,x') = g^{ab}
e^{\m'}_b(x',x')$ $ \na_{\m'} \s(x,x')$, $\s(x,x')$ being the geodetic interval
defined as one half the square of the length of this geodesic.
Then the generators of isometries can be presented in the form [20]
$$
\eqalignno{
P_a&=e^\m_{\ b}\left(\cos\sqrt K\right)^{b}_{\ a}\na_\m &(8.18)\cr
L_i&=e^\m_{\ b}\left({\sin\sqrt K\over \sqrt K}\right)^{b}_{\ a}
D^a_{\ ic}\s^c\na_\m. &(8.19)\cr}
$$
The functions of the matrix $K$ should be understood here as a power series in
curvature.

One can show that this operators form exactly the Lie algebra $\h G$ generated
by the curvature tensor of the symmetric space
$$
[\xi_A,\xi_B]=C^C_{\ AB}\xi_C,    \eqno(8.20)
$$
or, more explicitly,
$$
\eqalignno{
[P_a, P_b] &= E^i_{\ a b} L_i, \qquad &\cr
[P_a, L_i] &= D^b_{\  a i} P_b, \qquad &(8.21)\cr
[L_i, L_k] &= F^j_{\ i k} L_j, & \cr}
$$

Therefore, the curvature tensor of the symmetric space completely determines
the structure of the group of isometries.

Let us consider now the scalar case with {\it vanishing bundle curvature}
$$
\h R_{\m\n}=0. \eqno(8.21a)
$$
It is not difficult to show that in this case the Laplacian in symmetric space
can be presented in terms of generators of isometries
$$
\sq = g^{\m\n}\na_\m\na_\n = \g^{AB}\xi_A\xi_B =g^{ab}P_a P_b
+ \b^{ik}L_i L_k,         \eqno(8.22)
$$
where $\g^{AB}=(\g_{AB})^{-1}$ and $\b^{ik}=(\b_{ik})^{-1}$.

Using this representation one can prove a theorem expressing the heat kernel
operator in terms of some average over the group of isometries $G$
$$
\eqalignno{
\exp(t\sq) = &(4\pi t)^{-D/2} \int d k \g^{1/2}
	\det\left({\sinh(k^AC_A/2)\over k^AC_A/2}\right)^{1/2} &\cr
	& \times\exp\left\{ -{1\over 4t}k^A\g_{AB}k^B
	+ {1\over 6} R_G t\right\}\exp(k^A\xi_A) &(8.23)\cr}
$$
where $\g=\det\g_{AB}$ and $R_G$ is the scalar curvature of the group of
isometries $G$
$$
R_G= -{1\over 4}\g^{AB} C^C_{\ AD}C^D_{\ BC}. \eqno(8.24)
$$
The proof of this theorem is given in [20].

For further use it is convenient to rewrite the integral (8.23)
splitting the integration variables $k^A = (q^a, \om^i)$ in the form
$$
\eqalignno{
\exp&(t\sq) = (4\pi t)^{-D/2} \int dq\,d\om \eta^{1/2}\b^{1/2}
\det\left({\sinh((q^a C_a + \om^i C_i)/2)\over (q^a C_a + \om^i C_i)/2}
\right)^{1/2} &\cr
& \times\exp\left\{ -{1\over 4 t}(q^a g_{ab}q^b + \om^i\b_{ik}\om^k)
+ \left({1\over 8} R + {1\over 6} R_H \right) t \right\}
\exp\left(q^a P_a + \om^i L_i\right) ,  &\cr
&                 &(8.25)\cr}
$$
where$\b=\det \b_{ik}$, $\eta=\det g_{ab}$ and $R_H$ is the scalar curvature of
the isotropy subgroup $H$
$$
R_H = -{1\over 4} \b^{ik} F^m_{\ \ il}F^l_{\ km}. \eqno(8.26)
$$


To get the heat kernel in coordinate representation one has to act with the
heat kernel operator $\exp(t\sq)$ on the coordinate $\d$-function. Below we
will calculate only the heat kernel diagonal. Therefore, it is sufficient to
compute only the coincidence limit $x=x'$. Using the explicit form of the
generators of the isometries (8.18),(8.19) and solving the equations of
characteristics one can obtain the action of the isometries on the
$\d$-function [20]
$$
\exp\left(q^a P_a + \om^i L_i\right)g^{-1/2}\d(x,x')\Big\vert_{x=x'}
=\det\left({\sinh(\om^iD_i/2)\over \om^iD_i/2}\right)^{-1}\eta^{-1/2}\d(q).
\eqno(8.27)
$$

Now using (8.27) one can easily integrate over $q$ in (8.25) to get heat kernel
diagonal
$$
\eqalignno{
[U(t)]= &(4\pi t)^{-D/2}\int d\om \b^{1/2}
\det\left({\sinh(\om^iF_i/2)\over \om^iF_i/2}\right)^{1/2}
\det\left({\sinh(\om^iD_i/2)\over \om^iD_i/2}\right)^{-1/2} &\cr
&\times\exp\left\{ - {1\over 4 t}\om^i\b_{ik}\om^k
- \left(m^2+Q - {1\over 8} R
- {1\over 6} R_H \right) t \right\}.                     &(8.28)\cr}
$$

Changing the integration variables $\om \to \sqrt t \om$ and introducing a
Gaussian averaging over $\om$
$$
<f(\om)> = (4\pi)^{-p/2}\int d \om \b^{1/2}
\exp\left(-{1\over 4}\om^i\b_{ik}\om^k \right) f(\om)
							\eqno(8.29)
$$
one obtains then for the generating function $\Omega(t)$
we get another form of this formula
$$
\eqalignno{
\Omega&(t)=\exp\left\{\left({1\over 8}R
+{1\over 6}R_H\right)t\right\}& \cr
&\times\Bigg<\det\left({\sinh(\sqrt t \om^iF_i/2)\over \sqrt t
\om^iF_i/2}\right)^{1/2}
\det\left({\sinh(\sqrt t \om^iD_i/2)\over \sqrt t
\om^iD_i/2}\right)^{-1/2}\Bigg> &(8.30)\cr}
$$

One can present this result also in an alternative nontrivial rather {\it
formal} way. Substituting the equation
$$
(4\pi t)^{-p/2}\b^{1/2}\exp\left(-{1\over 4t} \om^i\b_{ik}\om^k\right) =
(2\pi)^{-p}\int dp \exp \left(ip_k\om^k -tp_k\b^{kn}p_n\right) \eqno(8.31)
$$
into the integral (8.28), integrating over $\om$ and changing the integration
variables $p_k \to i t^{-1/2} p_k $ we get finally an expression without any
integration
$$
\eqalignno{
\Omega(t)&=\exp\left\{\left({1\over 8}R
+{1\over 6}R_H\right)t\right\}\det\left({\sinh(\sqrt t \partial^kF_k/2)\over
\sqrt t \partial^kF_k/2}\right)^{1/2}&\cr
&\times
\det\left({\sinh(\sqrt t \partial^kD_k/2)\over \sqrt t
\partial^kD_k/2}\right)^{-1/2}
\exp\left(p_n\b^{nk}p_k\right)\Bigg\vert_{p=0}. &\cr
&       &(8.32)\cr}
$$
where $\partial^k=\partial/\partial p_k$.

This formal solution should be understood as a power series in the derivatives
$\partial^i$ that is well defined and determines the heat kernel asymptotic
expansion at $t\to 0$, i.e. {\it all} HMDS-coefficients $a_k$.

Let us mention that the formulae (8.30), (8.32) obtained in this section are
exact (up to possible nonanalytic topological contributions) and {\it
manifestly covariant} because they are expressed in terms of the invariants of
the isotropy group $H$, i.e. the invariants of the curvature tensor. They can
be used now to generate {\it all} HMDS-coefficients $a_k$ for {\it any}
symmetric space, i.e. for {\it any space with covariantly constant curvature},
simply by expanding it in a power series in $t$. Thereby one finds {\it all
covariantly constant terms in {\it all} HMDS-coefficients} in manifestly
covariant way. This gives a very nontrivial example how the heat kernel can be
constructed using only the commutation relations of some differential
operators, namely the generators of infinitesimal isometries of the symmetric
space. We are going to obtain the explicit formulae in a further work.

We considered for simplicity the case of symmetric space of {\it compact} type,
i.e. with positive sectional curvatures
$$
K(u,v)=R_{abcd}u^av^bu^cv^d=\b_{ik}(E^i_{\ ab}u^av^b)(E^k_{\ cd}u^cv^d),
\eqno(8.33)
$$
i.e. positive definite matrix $\b_{ik}$.
A simply connected symmetric space is, in general, reducible, and has the
following general structure [25]
$$
M=M_0\times M_+ \times M_- \eqno(8.34)
$$
where $M_0$, $M_+$ and $M_-$ are the {\it Euclidean, compact} and {\it
noncompact} components.

The corresponding algebra of isometries is a direct sum of ideals
$$
\h G = \h G_0\oplus\h G_+\oplus\h G_-      \eqno(8.35)
$$
where $\h G_0$ is an Abelian ideal and $\h G_+$ and ${\cal G}_-$ are the
semi-simple compact and noncompact ones.

There is a remarkable duality relation $*$ between compact and noncompact
objects. For any algebra $\h G = \h M + \h H = \{ P_a, L_i\} $ one defines the
dual one according to $\h G^* = i\h M + \h H = \{i P_a, L_k\}$, the structure
constants of the dual algebra being
$$
\{C^{*A}_{\ \ BC}\}=\{E^i_{\ ab}, D^c_{\ dk}, F^j_{\ lm}\}^*=\{ - E^i_{\ ab},
D^c_{\ dk}, F^j_{\ lm}\}.
							\eqno(8.36)
$$
So, the star $*$ only changes the sign of $E^i_{\ ab}$ but does not act
on all other structure constants. This means also that the matrix $\g$ (3.19)
for dual algebra should have the form
$$
\g^{*}_{AB} = \left(\matrix{ g_{ab} & 0             \cr
	       0                & \b_{ik}        \cr}\right)^*
=\left(\matrix{ g_{ab} & 0             \cr
	       0                & -\b_{ik}        \cr}\right) \eqno(8.37)
$$
and, therefore, the curvature of the dual manifold has the opposite sign
$$
R^*_{abcd}=-R_{abcd}. \eqno(8.38)
$$

We hope that it is not difficult to generalize our results to the general case
using the duality relation and the {\it analytic continuation}. This means that
our formulae (8.28), (8.30), (8.32) should be valid in general case of
arbitrary symmetric space too. Moreover, they should also be valid for the case
of pseudo-Euclidean signature of the metric $g_{\m\n}$.


\bigskip
\bigskip
\centerline{\mittel 9. Conclusion}
\bigskip

In present paper we have presented a brief overview of recent results in
studying the heat kernel obtained in our papers [20,21]. Here we have discussed
some ideas connected with the point that was
left aside in previous investigations [4,10,15-17], namely, the problem of
calculating the
low-energy limit of the heat kernel and the effective action in quantum gravity
and gauge theories.
We have analyzed in detail  the status of the low-energy approximation and
stressed the central role of an algebraic structure that naturally appears when
generalizing consistently the low-energy limit to curved space and gauge
theories. We have proposed a promising new purely algebraic approach for
calculating the low-energy heat kernel and realized, thereby, the idea of
partial summation of the terms without covariant derivatives in local
Schwinger - De Witt asymptotic expansion for computing the effective action
that was suggested in
[2,4].

Of course, there are left many unsolved problems. First of all, one has to
obtain {\it explicitly} the covariantly constant terms in HMDS-coefficients.
This would be the opposite case to the high-derivative approximation [15-17]
and can be of certain interest in mathematical phy\-sics.

Then, we still do not know how to calculate the low-energy heat kernel in
general case of covariantly
constant curvatures, i.e. when {\it all} background curvatures
($\Re=\{R_{\m\n\a\b}, {\cal R}_{\m\n}, Q \}$) are present. If it would be
possible to obtain in this general case the formulae
similar to (8.30) then it would allow to find the heat kernel in a number of
important cases and would lead finally to the general solution of the effective
potential problem in quantum gravity and gauge theories.

Besides, it is not perfectly clear how to do the analytical continuation of
Euclidean  low-energy effective action to the
space of Lorentzian signature for obtaining physical results.

\bigskip
\bigskip
\centerline{\mittel Acknowledgments }
\bigskip

I would like to express my sincere appreciation to the Organizing Committee of
the Conference `Heat Kernel Techniques and Quantum Gravity' for their kind
invitation and for the hospitality extended to me at the University of
Manitoba.


\bigskip
\bigskip

\centerline{\mittel References}
\bigskip

\item{[1]}  B. S. De Witt,
		{\it Dynamical theory of groups and fields} (Gordon and Breach,
		New York, 1965);
\item{}  B. S. De Witt, in: {\it Relativity, groups and topology II}, ed. by B.
S.  De Witt
		and R. Stora (North Holland, Amsterdam, 1984) p. 393
\item{[2]} G. A. Vilkovisky,
		in: {\it Quantum theory of gravity}, ed.  S. Christensen (Hilger,
		Bristol, 1983) p. 169
\item{[3]} A. O. Barvinsky and G. A. Vilkovisky,
		Phys. Rep. C 119 (1985) 1
\item{[4]} I. G. Avramidi,
		Nucl. Phys. B 355 (1991) 712
\item{} I. G. Avramidi, {\it The covariant methods for calculation of the
effective action in quantum field theory and the investigation of higher
derivative quantum gravity}, PhD Thesis (Moscow State
           University, Moscow, 1987)
\item{[5]} P. B. Gilkey,
		{\it Invariance theory, the heat equation and the  Atiyah - Singer
		index theorem} (Publish or Perish, Wilmington, DE, USA, 1984)
\item{[6]} J. Hadamard,
		{\it Lectures on Cauchy's Problem}, in: {\it Linear Partial Differential
		Equations} (Yale U. P., New Haven, 1923)
\item{} S. Minakshisundaram and A. Pleijel,
		Can. J. Math. 1 (1949) 242
\item{} R. T. Seely,
		Proc. Symp. Pure Math. 10 (1967) 288
\item{} H. Widom,
		Bull. Sci. Math. 104 (1980) 19
\item{} R. Schimming,
		Beitr. Anal. 15 (1981) 77
\item{}	R. Schimming,	Math. Nachr. 148 (1990) 145
\item{[7]} S. A. Fulling,
		SIAM J. Math. Anal. 13 (1982) 891
\item{} S. A. Fulling,		J. Symb. Comput. 9 (1990) 73
\item{}     S. A. Fulling and G. Kennedy,
		Trans. Am. Math. Soc. 310 (1988) 583
\item{[8]}  V. P. Gusynin,
		Phys. Lett. B 255 (1989) 233
\item{[9]} P. B. Gilkey,
		J. Diff. Geom. 10 (1975) 601
\item{[10]} I. G. Avramidi,
		Teor. Mat. Fiz. 79 (1989) 219
\item{} I. G. Avramidi,		Phys. Lett. B 238 (1990) 92
\item{[11]} P. Amsterdamski, A. L. Berkin and D. J. O'Connor,
		Class. Quantum  Grav. 6 (1989) 1981
\item{[12]} T. P. Branson  and P. B. Gilkey,
		Comm. Part. Diff. Eq. 15 (1990) 245
\item{}     N. Blazic, N. Bokan and P. B. Gilkey, Indian J. Pure Appl. Math.
		23 (1992) 103
\item{}     M. van den Berg and P. B. Gilkey, {\it Heat content asymptotics of
a
		Riemannian manifold with boundary}, University of Oregon
		preprint (1992)
\item{}     S. Desjardins and P. B. Gilkey, {\it Heat content asymptotics for
		operators of Laplace type with Neumann boundary conditions},
		University of Oregon preprint (1992)
\item{}     M. van den Berg, S. Desjardins and P. B. Gilkey, {\it Functorality
and
		heat content asymptotics for operators of Laplace type},
		University of Oregon preprint (1992)
\item{[13]}         G. Cognola, L. Vanzo and S. Zerbini,
		Phys. Lett. B 241 (1990) 381
\item{}     D. M. Mc Avity and H. Osborn,
		Class. Quantum Grav. 8 (1991) 603
\item{}	D. M. Mc Avity and H. Osborn,	Class. Quantum Grav. 8 (1991) 1445
\item{}	D. M. Mc Avity and H. Osborn,	Nucl. Phys. B 394 (1993) 728
\item{}     A. Dettki and A. Wipf,
		Nucl. Phys. B 377 (1992) 252
\item{}     I. G. Avramidi,
		Yad. Fiz. 56 (1993) 245
\item{[14]} P. B. Gilkey,
		{\it Functorality and heat equation asymptotics}, in: Colloquia
		Mathematica Societatis Janos Bolyai, 56. Differential
		Geometry, (Eger (Hungary), 1989), (North-Holland, Amsterdam,
		1992), p.  285
\item{} R. Schimming,
		{\it Calculation of the heat kernel coefficients}, in: B. Riemann
		Memorial Volume, ed. T. M.  Rassias, (World Scientific,
		Singapore), to be published

\item{[15]} I. G. Avramidi,
		Yad. Fiz. 49 (1989) 1185
\item{} I. G. Avramidi, Phys. Lett. B 236 (1990) 443
\item{[16]}     T. Branson, P. B. Gilkey and B. \O rsted,
		Proc. Amer. Math. Soc. 109 (1990) 437
\item{[17]} A. O. Barvinsky and G. A. Vilkovisky,
		Nucl. Phys. B 282 (1987) 163
\item{} A. O. Barvinsky and G. A. Vilkovisky, Nucl. Phys. B 333 (1990) 471
\item{}     G. A. Vilkovisky,
		{\it Heat kernel: recontre entre physiciens et mathematiciens},
		preprint CERN-TH.6392/92 (1992), in: Proc. of Strasbourg
		Meeting between physicists and  mathematicians (Publication de
		l' Institut de Recherche Math\'ematique  Avanc\'ee,
		Universit\'e  Louis
		Pasteur, R.C.P. 25, vol.43 (Strasbourg, 1992)), p. 203

\item{}     A. O. Barvinsky, Yu. V. Gusev, V. V. Zhytnikov and G. A.
		Vilkovisky, {\it Covariant perturbation theory (IY)}, Report
		of the University of Manitoba (University of Manitoba,
		Winnipeg, 1993)
\item{[18]} F. H. Molzahn, T. A. Osborn and S. A. Fulling,
		Ann. Phys. (USA) 204 (1990) 64
\item{} V. P. Gusynin, E. V. Gorbar and V. V. Romankov, Nucl Phys. B362 (1991)
449
\item{} P. B. Gilkey, T. P. Branson and S. A. Fulling,
		J. Math. Phys. 32 (1991) 2089
\item{} T. P. Branson, P. B. Gilkey and A. Pierzchalski,
		{\it Heat equation asymptotics of elliptic operators with
		non-scalar leading symbol}, University of Oregon preprint
		(1992)
\item{} F. H. Molzahn and T. A. Osborn,
		{\it A phase space fluctuation method for quantum dynamics},
		University of Manitoba preprint MANIT-93-01

\item{[19]} V. P. Gusynin,
		Nucl. Phys. B 333 (1990) 296
\item{} P. A. Carinhas and S. A. Fulling,
		in: {\it Asymptotic and computational  analysis}, Proc. Conf. in
		 Honor of Frank W. J. Olver's 65th birthday, ed.  R.  Wong
		(Marcel Dekker, New York, 1990) p. 601
\item{[20]} I. G. Avramidi,
		{\it Covariant methods for calculating the low-energy effective
		action in quantum field theory and quantum gravity},
		University of Greifswald (1994), gr-qc 9403036, submitted to J. Math. Phys.
\item{} I. G. Avramidi,
		{\it A new algebraic approach for calculating the heat kernel in quantum
gravity},
		University of Greifswald (1994), hep-th/9406047, submitted to Nucl. Phys. B
\item{} I. G. Avramidi,
		{\it The heat kernel on symmetric spaces via integrating over the group of
isometries}, University of Greifswald (1994), submitted to Phys. Lett. B

\item{[21]} I. G. Avramidi,
		Phys. Lett. B 305 (1993) 27;

\item{[22]} J. S. Dowker,
		Ann. Phys. (USA) 62 (1971) 361
\item{} J. S. Dowker,	J. Phys. A 3 (1970) 451
\item{} A. Anderson and R. Camporesi,
		Commun. Math. Phys. 130 (1990) 61
\item{} R.  Camporesi,
		Phys. Rep. 196 (1990) 1
\item{} N. E. Hurt,
		{\it Geometric quantization in action:  applications of harmonic
		analysis in quantum statistical mechanics and quantum field
		theory}, (D. Reidel Publishing Company,  Dordrecht, Holland,
		1983)

\item{[23]} E. S. Fradkin and A. A. Tseytlin,
		Nucl. Phys. B 234 (1984) 472
\item{}  I. G. Avramidi,
		{\it Covariant algebraic calculation of the one-loop effective potential in
non-Abelian gauge theory and a new approach to stability problem}, University
of Greifswald (1994), gr-qc 9403035, submitted to J. Math. Phys.
\item{} I. L. Buchbinder, S. D. Odintsov, I. L. Shapiro, {\it Effective action
in quantum gravity} (IOP Publishing,
		Bristol, 1992)
\item{} G. Cognola, K. Kirsten and S. Zerbini,
		{\it One-loop effective potential on hyperbolic manifolds},
		Trento University preprint (1993)
\item{} A. Bytsenko, K. Kirsten and S. Odintsov,
		{\it Self-interacting scalar fields on spacetime with compact
		hyperbolic spatial part}, Trento University preprint (1993)
\item{[24]} J. A. Zuck,
		Phys. Rev. D 33 (1986) 3645
\item{}     V. P. Gusynin and V. A. Kushnir,
		Class. Quantum Grav. 8 (1991) 279

\item{[25]} H. S. Ruse, A. G. Walker, T. J. Willmore, {\it Harmonic spaces},
                (Edizioni Cremonese, Roma (1961))
\item{} J. A. Wolf, {\it Spaces of constant curvature} (University of
		California, Berkeley, CA, 1972)
\item{}  B. F. Dubrovin, A. T. Fomenko and S. P. Novikov,
		{\it The Modern  Geometry: Methods and  Applications} (Springer, N.Y.
		1992)

\bye